\documentclass[a4paper]{article}
\title{
\vspace{-3mm}
\rightline{\small IFUP-TH 2001/29}
\vspace{8mm}
\bf Finite Temperature Behavior of the\\
3D Polyakov Model with Massless Quarks}
\author{Nikita O. Agasian \thanks{
E-mail address: {\tt agasian@heron.itep.ru}}
\\
{\it Institute of Theoretical and Experimental Physics,}\\
{\it B. Cheremushkinskaya 25, RU-117 218 Moscow, Russia}\\
and\\
Dmitri Antonov \thanks{
E-mail address: {\tt antonov@df.unipi.it}}
\thanks{Permanent address:
ITEP, B. Cheremushkinskaya 25, RU-117 218 Moscow, Russia.}\\
{\it INFN-Sezione di Pisa, Universit\'a degli studi di Pisa,
Dipartimento di Fisica,}\\
{\it Via Buonarroti, 2 - Ed. B -
I-56127 Pisa, Italy}}

\date{}
\begin{document}

\maketitle
\vspace{1mm}
\centerline{\bf {Abstract}}
\vspace{3mm}
\noindent
The (2+1)D Georgi-Glashow (or Polyakov)
model with the additional fundamental massless quarks
is explored at finite temperature.
In the case of vanishing Yukawa coupling,
it is demonstrated that the interaction of
a monopole and an antimonopole in the molecule
via quark zero modes leads to the decrease of the
Berezinsky-Kosterlitz-Thouless critical temperature when the
number of quark flavors is equal to one. If the number of flavors
becomes larger, monopoles are shown to exist
only in the molecular phase at any temperatures exceeding a certain
exponentially small one.
This means that
for such a number of flavors and at such temperatures, no fundamental matter
can be confined by means of the monopole mechanism.

\vspace{10mm}

3D Georgi-Glashow model (else called the Polyakov model) is known to be
one of the eldest and the most famous examples of
theories allowing for an analytical description
of confinement~\cite{Polyakov}.
However, the phase structure of this model at finite temperature
has been addressed only recently. Namely, first in Ref.~\cite{1} it
has been shown that at the temperature $T_c=g^2/ 2\pi$ the
weakly coupled monopole plasma in this model
undergoes the Berezinsky-Kosterlitz-Thouless (BKT)~\cite{bkt}
phase transition into the molecular phase. Then, in
Ref.~\cite{2}, it has been shown that approximately
at the twice smaller temperature,
there occurs another phase transition associated to the
deconfinement of W-bosons.

In this paper, we shall
be interested in the finite-temperature
properties of the monopole ensemble, rather than the ensemble of W-bosons.
Because of that, let us first discuss in some more details the nature of the
above-mentioned BKT phase transition.
At high enough temperature, one can apply the idea of dimensional
reduction.
The dimensionally-reduced theory is then the 2D XY-model,
but with the temperature-depending strength of the
monopole-antimonopole ($M \bar M$) interaction. Due to this fact, the phase structure
of the model becomes reversed with respect to that of the usual 2D XY-model. Namely, at
the temperatures below $T_c$, monopoles exist in the plasma phase, that leads to the
confinement of fundamental matter~\cite{Polyakov, 5}. At $T>T_c$, the vacuum state
is the molecular gas of bound $M \bar M-$pairs, and consequently
fundamental quarks are deconfined~\cite{1}. The analogy with the 2D XY-model
established in Ref.~\cite{bmk} is that spin waves of the 2D XY-model correspond
to the free photons of the Polyakov model, while vortices correspond to magnetic
monopoles.

Let us briefly discuss the BKT phase transition, occuring at $T=T_c$,
in the language of the 2D XY-model. At $T<T_c$, the spectrum of the model
is dominated by massless spin waves, and the periodicity of the angular
variable is unimportant in this phase. The spin waves are unable to disorder
the spin-spin correlation functions, and those decrease at large distances by
a certain power law. On the contrary, at $T>T_c$,
the periodicity of the angular variable becomes important. This leads to the
appearance of topological singularities (vortices) of the angular variable,
which, contrary to spin waves, have nonvanishing winding numbers.
Such vortices condense and disorder the spin-spin correlation
functions, so that those start decreasing exponentially with the distance.
Thus, the nature of the BKT phase transition is the condensation
of vortices at $T>T_c$. In another words, at $T>T_c$, there exist free vortices,
which mix in the ground state (vortex condensate) of indefinite
global vorticity. Contrary to that, at $T<T_c$, free vortices cannot exist,
but they rather mutually couple into bound states of vortex-antivortex pairs.
Such vortex-antivortex molecules are small-sized short-living (virtual)
objects. Their dipole-type fields are short-ranged and therefore cannot
disorder significantly the spin-spin correlation functions. However, when
the temperature starts rising, the sizes of these molecules increase,
until at $T=T_c$ they diverge, that corresponds to the dissociation of the molecules
into pairs. Therefore, coming back to the Polyakov model,
one of the methods (which will be employed below) to determine there the critical
temperature of the BKT phase transition is to evaluate the mean squared separation
in the $M \bar M-$molecule and find the temperature at which it
starts diverging.

In this paper, we shall consider the extension of the Polyakov model by
the fundamental dynamical quarks, which are supposed to be massless.
As it will be demonstrated,
quark zero modes in the monopole field lead to the additional attraction
between a monopole and an antimonopole in the molecule at high temperatures.
In particular, when the number of these modes (equal to the
number of massless flavors) is sufficiently
large, the molecule shrinks so that its size becomes of the order
of the inverse W-boson mass. Another factor which governs the size of the
molecule is the characteristic range of localization of zero modes. Namely, it can be shown that
the stronger zero modes are localized in the vicinity of the monopole center, the
smaller molecular size is. In this paper,
we shall consider the case when the Yukawa coupling vanishes,
and originally massless quarks do not acquire any mass. This means that
zero modes are maximally delocalized.
Such a weakness of the quark-mediated interaction of monopoles
opens a possibility for molecules to
undergo eventually the phase transition
into the plasma phase.
However, this will be shown to occur only provided that
the number of flavors is equal to one, whereas at any larger
number of flavors,
the respective critical temperature becomes exponentially small.
This means that the interaction mediated by such a number
of zero modes is already strong enough to maintain the molecular phase
at any temperature larger than that one.

Let us start our analysis with considering the
Euclidean action of the Polyakov model extended by the fundamental,
originally massless quarks.
[Note that such a model can be viewed as the (2+1)-QCD with the additional
adjoint Higgs field.]
We shall first consider the general case with the nonvanishing
Yukawa coupling, by means of which quarks acquire a certain mass.
The action under discussion then reads

\begin{equation}
\label{GG}
S=\int d^3x\left[\frac{1}{4}\left(F_{\mu\nu}^a\right)^2+
\frac12\left(D_\mu\Phi^a\right)^2+\frac{\lambda}{4}\left(
\left(\Phi^a\right)^2-\eta^2\right)^2-i\bar\psi\left(\vec\gamma
\vec D+h\frac{\tau^a}{2}\Phi^a\right)\psi\right].
\end{equation}
Here,

$$F_{\mu\nu}^a=\partial_\mu A_\nu^a-\partial_\nu A_\mu^a+
g\varepsilon^{abc}A_\mu^bA_\nu^c,~
D_\mu\Phi^a=\partial_\mu\Phi^a+g\varepsilon^{abc}A_\mu^b
\Phi^c,~
D_\mu\psi=\left(\partial_\mu-ig\frac{\tau^a}{2}A_\mu^a
\right)\psi,$$
and $\bar\psi=\psi^{\dag}\beta$ with
the Euclidean Dirac matrices defined as $\vec\gamma=
-i\beta\vec\alpha$, where

$$\beta=\left(
\begin{array}{cc}
1& 0\\
0& -1
\end{array}
\right),~~
\vec\alpha=\left(
\begin{array}{cc}
0& \vec\tau\\
\vec\tau& 0
\end{array}
\right).
$$
Next, in 3D, the electric coupling $g$, the Yukawa
coupling $h$, and the vacuum expectation value of the Higgs field $\eta$
have the dimensionality $[{\rm mass}]^{1/2}$. The Higgs coupling $\lambda$
has the dimensionality $[{\rm mass}]$.
The masses of the W- and Higgs bosons are large compared to $g^2$
in the standard perturbative (else called weak-coupling)
regime $g\ll\eta$ and read: $m_W=g\eta$, $m_H=\eta\sqrt{2\lambda}$.
The inequality $g\ll\eta$ is necessary to ensure the spontaneous
symmetry breaking from $SU(2)$ to $U(1)$.
Note also that for the sake of simplicity,
we omit the summation over the flavor indices, but consider the
general case with an arbitrary number of flavors.

One can further see that the Dirac equation in the field of the third
isotopic component of the 't Hooft-Polyakov monopole~\cite{3} decomposes
into two equations for the components of the $SU(2)$-doublet $\psi$.
The masses of these components stemming from such equations are
equal to each other and read $m_q=h\eta/2$.
Next, the Dirac equation in the full monopole potential has been
shown~\cite{4} to possess the zero mode, whose
asymptotic behavior at $r\equiv\left|\vec x{\,}\right|\gg m_q^{-1}$
has the following form:

\begin{equation}
\label{N}
\chi_{\nu{\,}n}^{+}={\cal N}\frac{{\rm e}^{-m_qr}}{r}
\left(s_\nu^{+}s_n^{-}-s_\nu^{-}s_n^{+}\right),~~
\chi_{\nu{\,}n}^{-}=0.
\end{equation}
Here, $\chi^{\pm}_n$ are the upper and the lower components of the mode,
{\it i.e.} $\psi={\chi_n^{+}\choose \chi_n^{-}}$, next
$n=1,2$ is the isotopic index, $\nu=1,2$ is the Dirac index,
$s^{+}={1\choose 0}$, $s^{-}={0\choose 1}$, and ${\cal N}$
is the normalization constant.

It is a well known fact that in 3D, the 't Hooft-Polyakov monopole is actually
an instanton~\cite{Polyakov, 5}. Therefore, we can use the results of Ref.~\cite{6} on the quark
contribution to the effective action of the instanton-antiinstanton molecule
in QCD. Let us thus recapitulate
the analysis of Ref.~\cite{6} adapting it to our model.
To this end, we fix the gauge $\Phi^a=\eta\delta^{a3}$ and
define the analogue of the free propagator $S_0$ by the relation
$S_0^{-1}=-i\left(\vec\gamma\vec\partial+m_q\tau^3\right)$.
Next, we define the
propagator $S_M$ in the field of a monopole located at the origin,
$\vec A^{a{\,}M}$
[$A_i^{a{\,}M}\to\varepsilon^{aij}x^j/\left(gr^2\right)$ at $r\gg m_W^{-1}$],
by the formula $S_M^{-1}=S_0^{-1}-g\vec\gamma\frac{\tau^a}{2}\vec A^{a{\,}M}$.
Obviously, the propagator $S_{\bar M}$ in the field of an antimonopole
located at a certain point $\vec R$,
$\vec A^{a{\,}\bar M}\left(\vec x\right)=-\vec A^{a{\,}M}\left(\vec x-\vec R
\right)$, is defined by the equation for
$S_M^{-1}$ with the replacement $\vec A^{a{\,}M}\to \vec A^{a{\,}\bar M}$.
Finally, one can consider the molecule made out of these monopole
and antimonopole and define the total propagator $S$ in the field of such a
molecule, $\vec A^a=\vec A^{a{\,}M}+\vec A^{a{\,}\bar M}$,
by means of the equation for $S_M^{-1}$ with $\vec A^{a{\,}M}$
replaced by $\vec A^a$.

One can further
introduce the notation $\left|\psi_n\right>$, $n=0,1,2,\ldots$,
for the eigenfunctions of the operator $-i\vec\gamma\vec D$ defined
at the field of the molecule, namely $-i\vec\gamma\vec D
\left|\psi_n\right>=\lambda_n\left|\psi_n\right>$, where
$\lambda_0= 0$. This yields the following formal spectral
representation for the total propagator $S$:

$$S\left(\vec x, \vec y\right)=\sum\limits_{n=0}^{\infty}
\frac{\left|\psi_n(\vec x)\right>\left<\psi_n(\vec y)\right|}
{\lambda_n-im_q\tau^3}.$$
Next, it is convenient to employ the mean-field approximation,
according to which zero modes dominate in the quark propagator, {\it i.e.},

\begin{equation}
\label{domin}
S\left(\vec x, \vec y\right)\simeq
\frac{\left|\psi_0(\vec x)\right>\left<\psi_0(\vec y)\right|}
{-im_q\tau^3}+S_0\left(\vec x, \vec y\right).
\end{equation}
Indeed, this approximation is valid, since in the weak-coupling regime
the monopole sizes, equal to $m_W^{-1}$, are much smaller than
the average distance in the $M \bar M-$plasma.
This average distance has an order of
magnitude $\zeta^{-1/3}$
(see {\it e.g.} Ref.~\cite{9} for a discussion).
Here, $\zeta\propto{\rm e}^{-4\pi m_W \epsilon/g^2}$
stands for the so-called monopole fugacity,
which has the dimensionality $[{\rm mass}]^3$, and $\epsilon\sim 1$ is a
certain dimensionless function of $(m_H/m_W)$.
Obviously, $\zeta$ is
exponentially small in the weak-coupling regime under study.
The approximation~(\ref{domin}) remains valid
for the molecular phase near the phase transition ({\it i.e.} when the
temperature approaches the critical one from above), we shall be interested in.
That is merely because in this regime, molecules become very much
inflated being about to dissociate.

Within the notations adapted, one now has
$S=\left(S_M^{-1}+S_{\bar M}^{-1}-S_0^{-1}\right)^{-1}=
S_{\bar M}{\cal S}^{-1}S_M$, where

$${\cal S}= S_0-\left(S_M-S_0\right)S_0^{-1}
\left(S_{\bar M}-S_0\right)=S_0-
\frac{\Bigl|\psi_0^M\Bigr>\Bigl<\psi_0^M\Bigr|}{-im_q\tau^3}
S_0^{-1}\frac{\Bigl|\psi_0^{\bar M}\Bigr>
\Bigl<\psi_0^{\bar M}\Bigr|}{-im_q\tau^3},$$
and $\Bigl|\psi_0^M\Bigr>$, $\Bigl|\psi_0^{\bar M}\Bigr>$ are the
zero modes of the operator $-i\vec\gamma\vec D$ defined at the field of
a monopole and an antimonopole, respectively. Denoting further
$a=\left<\psi_0^{\bar M}\left|g\vec\gamma\frac{\tau^a}{2}
\vec A^{a{\,}M}\right|\psi_0^M\right>$,
it is straightforward to see by the definition of the zero mode
that $a=\left<\psi_0^{\bar M}\left|\left(-i\vec\gamma\vec\partial\right)
\right|\psi_0^M\right>=\left<\psi_0^{\bar M}\left|S_0^{-1}
\right|\psi_0^M\right>$. This yields ${\cal S}=S_0+({a^{*}}/{m_q^2})
\Bigl|\psi_0^M\Bigr>\Bigl<\psi_0^{\bar M}\Bigr|$, where the star
stands for the complex conjugation, and therefore $\det {\cal S}=
\left[1+\left({|a|}/{m_q}\right)^2\right]\cdot\det S_0$. Finally,
defining the desired effective action as $\Gamma=\ln
[{\det S^{-1}}/{\det S_0^{-1}}]$, we obtain for it in the
general case with $N_f$ flavors the following expression:
$\Gamma={\,}{\rm const}{\,}+N_f\ln\left(m_q^2+|a|^2\right)$.
The constant in this formula, standing for the
sum of effective actions defined at the monopole and at the antimonopole,
cancels out in the normalized expression for the mean squared
separation in the $M \bar M-$molecule.

Let us further set $h$ equal to zero, and so $m_q$ is equal to zero as well.
Notice first of all that although in this case the direct Yukawa
interaction of the Higgs bosons with quarks is absent, they keep
interacting with each other
via the gauge field. Owing to this fact, the problem of
finding a quark zero mode in the monopole field is still valid~\footnote{
Note that according to Eq.~(\ref{N}) this mode will be non-normalizable
in the sense of a discrete spectrum.
However, in the gapless case $m_q=0$ under discussion, the zero mode,
which lies exactly on the border of the two contiguous Dirac seas, should
clearly be treated not as an isolated state
of a discrete spectrum, but rather
as a state of a continuum spectrum. (A similar
treatment of the zero mode of a massless left-handed neutrino on
electroweak Z-strings has been discussed in Ref.~\cite{10}.)
 This means that it should be
understood as follows: $\left|\psi_0(\vec x)\right>
\sim\lim\limits_{p\to 0}^{}\left({\rm e}^{ipr}/r\right)$, where $p=
\left|\vec p{\,}\right|$. Once being considered in this way, zero modes
are normalizable by the standard condition of normalization of the
radial parts of spherical waves, $R_{pl}$, which reads~\cite{11}
$\int\limits_{0}^{\infty}drr^2R_{p'l}R_{pl}=2\pi\delta(p'-p)$.}.
The dependence of the
absolute value of the matrix element $a$ on the distance $R$ between
a monopole and an antimonopole
can now be straightforwardly found. Indeed, we have
$|a|\propto\int d^3r/\left(r^2\left|\vec r-\vec R\right|\right)=
-4\pi\ln(\mu R)$, where $\mu$ stands for the IR cutoff.

Now we switch on the temperature $T\equiv\beta^{-1}$,
so that all the bosonic (fermionic) fields should be supplied
with the periodic (antiperiodic) boundary conditions in the
temporal direction, with the period equal to $\beta$. The magnetic-field
lines of a single monopole thus cannot cross the boundary of the
one-period region and should go parallel
to this boundary at the distances larger than $\beta$. Therefore,
monopoles separated by such distances interact via the 2D Coulomb
law, rather than the 3D one. Recalling that the average distance
between monopoles is of the order of $\zeta^{-1/3}$, we conclude
that at $T\ge\zeta^{1/3}$, the monopole ensemble becomes two-dimensional
(see {\it e.g.} Ref.~\cite{1} for a detailed discussion
of the dimensional reduction in the Polyakov model).
However, at the temperatures
below the exponentially small one, $\zeta^{1/3}$, monopoles
keep interacting by the usual 3D Coulomb law, and the monopole confinement
mechanism for the fundamental matter works at such temperatures
under any circumstances.

We are now in the position to explore a possible modification of the
standard BKT critical temperature~\cite{1} $T_c=g^2/ 2\pi$
due to the zero-mode mediated interaction. As it was discussed above,
this can be done upon the evaluation of
the mean squared separation in the $M \bar M-$molecule
and further finding the temperature below which it starts diverging.
In this way we should take into account that in the dimensionally-reduced
theory, the usual Coulomb interaction of monopoles~\footnote{Without
the loss of generality, we consider the molecule with the temporal component
of $\vec R$ equal to zero.}
$R^{-1}=
\sum\limits_{n=-\infty}^{+\infty}\left({\cal R}^2+(\beta n)^2\right)^{-1/2}$
goes over into $-2T\ln(\mu{\cal R})$, where ${\cal R}$ denotes the
absolute value of the 2D vector $\vec{\cal R}$. This statement can be checked
{\it e.g.} by virtue of the Euler - Mac~Laurin formula.
As far as the novel logarithmic interaction,
$\ln(\mu R)= \sum\limits_{n=-\infty}^{+\infty}\ln\left[\mu
\left({\cal R}^2+(\beta n)^2\right)^{1/2}\right]$, is concerned, it
transforms into

\begin{equation}
\label{sumlog}
\pi T{\cal R}+\ln\left[1-\exp(-2\pi T{\cal R})\right]-\ln 2.
\end{equation}
Let us prove this statement.
To this end, we employ the following formula~\cite{gr}:

$$
\sum\limits_{n=1}^{\infty}\frac{1}{n^2+x^2}=\frac{1}{2x}
\left[\pi\coth(\pi x)-\frac{1}{x}\right].
$$
This yields

$$
x\sum\limits_{n=-\infty}^{+\infty}\frac{1}{x^2+\left({2\pi n}/{a}\right)^2}=
\frac{1}{x}+\frac{xa^2}{2\pi^2}\sum\limits_{n=1}^{\infty}\frac{1}{n^2+
\left({xa}/{2\pi}\right)^2}=\frac{a}{2}\coth\left(\frac{ax}{2}\right).
$$
On the other hand, the L.H.S. of this expression can be written as

$$\frac{1}{2}
\frac{d}{dx}\sum\limits_{n=-\infty}^{+\infty}\ln\left(x^2+\left(\frac{2\pi n}{a}\right)^2
\right).
$$
Integrating over $x$ with the constant of integration set to zero, we get

$$
\sum\limits_{n=-\infty}^{+\infty}\ln\left(x^2+\left(\frac{2\pi n}{a}\right)^2
\right)=a\int dx\coth\left(\frac{ax}{2}\right)=2\ln\sinh\left(\frac{ax}{2}\right)=ax+
2\ln\left(1-{\rm e}^{-ax}\right)-2\ln 2.
$$
Setting ${2\pi}/{a}=\mu \beta$ and $x=\mu {\cal R} $ we arrive at Eq.~(\ref{sumlog}).

Thus, the statistical weight of the quark-mediated interaction in the
molecule at high temperatures has the form
$\exp({-2N_f\ln|a|})\propto\left[
\pi T{\cal R}+\ln\left[1-\exp(-2\pi T{\cal R})\right]-\ln 2
\right]^{-2N_f}$. Accounting for both (former) logarithmic and Coulomb
interactions, we eventually arrive at the following expression for the
mean squared separation $\left<L^2\right>$ in the molecule as a function of
$T$, $g$, and $N_f$:

$$
\left<L^2\right>=\frac{
\int\limits_{m_W^{-1}}^{\infty}d{\cal R}{\cal R}^{3-\frac{8\pi T}{g^2}}
\left[\pi T{\cal R}+\ln\left[1-\exp(-2\pi T{\cal R})\right]-\ln 2\right]^{-2N_f}}
{\int\limits_{m_W^{-1}}^{\infty}d{\cal R}{\cal R}^{1-\frac{8\pi T}{g^2}}
\left[\pi T{\cal R}+\ln\left[1-\exp(-2\pi T{\cal R})\right]-\ln 2\right]^{-2N_f}}.
$$
In this equation, we have put the lower limit of integration
equal to the inverse mass of the W-boson, which acts as an UV cutoff.

At large ${\cal R}$, $\ln 2 \ll \pi T{\cal R}$ and
$\bigl|\ln\left[1-\exp(-2\pi T{\cal R})\right]\bigr|\simeq
\exp(-2\pi T{\cal R})\ll
\pi T{\cal R}$. Consequently, we see that $\left<L^2\right>$ is finite at
$T>T_c=(2-N_f)g^2/ 4\pi$, that reproduces the standard result~\cite{1}
at $N_f=0$. For $N_f=1$, the plasma phase
is still present at $T<g^2/ 4\pi$,
whereas for $N_f\ge 2$ the monopole ensemble may exist only
in the molecular phase at any temperature larger than $\zeta^{1/3}$.
Clearly, at $N_f\gg \max\left\{1,{4\pi T}/{g^2}\right\}$,
$\sqrt{\left<L^2\right>}\to m_W^{-1}$,
which means that such a large number of zero modes shrinks the molecule
to the minimal admissible size.
Note finally that both the obtained critical temperature $g^2/ 4\pi$
and the standard one (in the absence of quarks), $g^2/ 2\pi$,
are obviously much larger than $\zeta^{1/3}$, that fully validates
the idea of dimensional reduction.

In conclusion of this paper, we have found the
critical temperature of the monopole BKT phase transition
in the weak-coupling regime of the Polyakov model
extended by the massless dynamical quarks, which interact with the Higgs boson only
via the gauge field. It has been shown that for $N_f=1$, this
temperature becomes twice smaller than the one in the
absence of quarks, whereas for $N_f\ge 2$ it becomes exponentially
small, namely of the order of $\zeta^{1/3}$. The latter effect
means that this number of quark zero modes, which strengthen the attraction
of a monopole and an antimonopole in the molecule, becomes enough
for the support of the molecular phase at any temperature exceeding that
exponentially small one. Therefore, for $N_f\ge 2$, no fundamental matter
(including dynamical quarks themselves) can be confined at
such temperatures by means of the monopole mechanism.

\section*{Acknowledgments}

We are greatful for stimulating and fruitful discussions to
A.~Di~Giacomo, B.O.~Kerbikov, I.I.~Kogan, V.A.~Rubakov, M.A.~Shifman,
and Yu.A.~Simonov. D.A. is indebted to Prof. A.~Di~Giacomo and the
whole staff of the Quantum Field Theory Division of the University
of Pisa for kind hospitality and to INFN for the financial
support. And last but not least, partial financial support by the
RFBR grant 00-02-17836 and the INTAS grant Open Call 2000, project
No. 110, is greatfully acknowledged.

%\newpage

\end{document}